\documentclass[reprint,showpacs,preprintnumbers,amsmath,amssymb, aip,jcp,nolongbibliography]{revtex4-1}
\usepackage{fancyhdr}
\usepackage[usenames,dvipsnames]{color}
\usepackage{graphicx}% Include figure files
\usepackage{dcolumn}% Align table columns on decimal point
\usepackage{bm}% bold math
 \usepackage{float}
\usepackage{natbib}
\begin{document}

\preprint{1}

\title{The Putative Liquid-Liquid Transition is a Liquid-Solid Transition in Atomistic Models of Water }

\author{David T. Limmer}
\author{David Chandler}
 \email{chandler@berkeley.edu}
\affiliation{%
Department of Chemistry, University of California, Berkeley, California 94720
}%

\date{\today}
\begin{abstract}
We use numerical simulation to examine the possibility of a reversible liquid-liquid transition in supercooled water and related systems. In particular, for two atomistic models of water, we have computed free energies as functions of multiple order parameters, where one is density and another distinguishes crystal from liquid.  For a range of temperatures and pressures, separate free energy basins for liquid and crystal are found, conditions of phase coexistence between these phases are demonstrated, and time scales for equilibration are determined. We find that at no range of temperatures and pressures is there more than a single liquid basin, even at conditions where amorphous behavior is unstable with respect to the crystal. We find a similar result for a related model of silicon. This result excludes the possibility of the proposed liquid-liquid critical point for the models we have studied. Further, we argue that behaviors others have attributed to a liquid-liquid transition in water and related systems are in fact reflections of transitions between liquid and crystal.  
\end{abstract}

\maketitle

\section{\label{sec:level1}Introduction}
This paper reports the results of a numerical study aimed at elucidating the purported \cite{Poole:1992p2103,Mishima:1998p2948} liquid-liquid phase transition in supercooled liquid water.  The results indicate that this hypothesized polyamorphism does not exist in atomistic models of water.  While not contradicting the existence of irreversible polyamorphism of the sort observed in non-equilibrium disordered solids of water,\cite{Johari:1987p5768,Hallbrucker:1989p5814,Angell:1995p3350, Poole:1997p2912,Loerting:2001p5935} and not excluding the possibilities of liquid-liquid transitions in liquid mixtures \cite{Sengers:1986p4788}, polymerizing fluids \cite{Katayama} and some theoretical models,\cite{Jagla:2001p4642,Wilding:2002p4638,Poole:1994p673,Stokely:2010p6455} the results do suggest that a reversible transition and its putative second critical point are untenable for one-component liquids, like water, that exhibit local tetrahedral order and freeze into crystals with similar but extended order. 

The terminology ``transition'' is used here to refer to distinct phases, where coexistence implies the formation of interfaces that would spatially separate the coexisting phases or to response functions that diverge in the thermodynamic limit.\cite{Chandler_book}  The structural changes for a transition between two liquids or between a liquid and a crystal are distinct from continuous pressure induced changes in normal liquid water.\cite{Soper:2000p6573} These changes associated with a phase transition are global and therefore are also distinct from bi-continuous behaviors that do not persist beyond small length scales.\cite{Pettersen_Paper}

Polyamorphism of water has been achieved through various out-of-equilibrium experimental protocols resulting in a multitude of thermodynamically unstable, kinetically trapped structures.\cite{Burton:1935p3483,Narten:1976p3715,Brueggeller:1980p3811,Mishima:1984p3884,Mishima:1985p2219,Mayer:1985p3950} These different disordered structures have been generally partitioned into two general categories known as either low-density amorphous solids \cite{Burton:1935p3483,Narten:1976p3715,Brueggeller:1980p3811} or high-density amorphous solids.\cite{Mishima:1984p3884,Mishima:1985p2219,Mayer:1985p3950} Some have interpreted changes in these structural motifs as non-equilibrium manifestations of an underlying equilibrium phase transition between two forms of liquid water. This conjecture forms the basis of some attempts to explain many of the well-known anomalous thermodynamic properties of water, e.g., Refs. \onlinecite{Poole:1992p2103}, \onlinecite{BellissentFunel:1998p3960}, \onlinecite{Fuentevilla:2006p4635} and \onlinecite{Kumar:2008p6553}. There are other ways of explaining these anomalies,\cite{Sastry:1996p2944} but the phase transition hypothesis seems particularly intriguing, and it is the focus of this paper.

The hypothesis is impossible to test by natural experiments because the location of the presumed transition is outside experimentally accessible conditions.\cite{Mishima:2000p4162} In particular, bulk supercooled water is unstable as a liquid, and it rapidly crystallizes in the regime of predicted polyamorphism.  While the properties of non-equilibrium glassy materials can be studied in this region, it is uncertain whether inferences regarding reversible thermodynamic behavior can be made from such measurements.  Some experiments have studied water confined to long pores with radii no larger than 1 nm.\cite{Liu:2005p2412,Liu:2007p2002,Mallamace:2007p1528,Zhang:2010p4637}  These experiments avoid the instability and thereby attempt to detect manifestations of the transition. While water does not freeze in such circumstances, it is questionable whether properties of bulk water can be inferred from behaviors of these one-dimensional systems.\cite{Findenegg:2008p1458,Mancinelli:2010p763} 

Molecular simulation provides a means to overcome this ambiguity.  Specifically, sufficiently realistic models can be studied computationally while controlling order parameters that distinguish liquid from crystal.  It is in this way that we examine the reversible behavior of models of water.  Along with establishing coexistence between liquid and crystal, we are able to study the dynamics of the transition between these phases.  We also locate and explore the free energy surface for the region of the pressure-temperature phase diagram known as ``no man's land''.\cite{Mishima:1998p2948}  This is the region where amorphous behavior would be unstable in the absence of control.  Our results indicate that some observations attributed by others as manifestations of a liquid-liquid transition are in fact observations of the temperature-pressure boundary separating the region of amorphous instability from that of a single phase of amorphous metastability. 

Others have used molecular simulation for realistic models of water \cite{Brovchenko:2003p206,Liu:2009p776,Moore:2010p1923,Kumar:2005p2095,Poole:2005p6308, Harrington:1997p202} and related liquids \cite{Sastry:2003p234,Beaucage:2005p4732,Widom:2009p4690,Vasisht:2011p6552} to examine their possible polyamorphisms.  Those cited here\cite{Harrington:1997p202,Brovchenko:2003p206,Liu:2009p776,Moore:2010p1923,Sastry:2003p234,Beaucage:2005p4732,Widom:2009p4690, Vasisht:2011p6552, Kumar:2005p2095,Poole:2005p6308} are representative but by no means comprehensive (we exclude from this list models that do not exhibit local tetrahedral order, e.g. Ref.~\onlinecite{Jagla:2001p4642}). In all cases, the methods employed have been limited in at least one of three ways: time scales that are too short, system sizes that are too small, and order parameters that fail to discriminate order from disorder or fail to be adequately controlled.  By employing multiple order parameters and free-energy sampling methods, we overcome time-scale issues and are able to discriminate between phases of different symmetries.   By considering different system sizes and size scaling analysis, we overcome uncertainty associated with finite system sizes.

Most of the results we present in this paper have been computed with a recently developed model by Molinero, so-called ``mW'' water.\cite{Molinero:2008p4576}  We use this model for three reasons.  First, it is a computationally convenient model because it contains no long-ranged forces, relying instead on short-ranged three-body forces to favor microscopic structures consistent with those of water.  Second, the behavior of the model is realistic in the sense that in the range of conditions we wish to study its phase diagram is a reasonable caricature of that for water.\cite{Jacobson:2009p2605,Moore:2010p1923,Molinero:2011p4959}  Third, the results obtained with this model would seem to apply to other systems in addition to water in that the model is a variant of one developed by Stillinger and Weber,\cite{Stillinger:1985p3216} which has been used to treat behaviors of Si \cite{Sastry:2003p234,Beaucage:2005p4732,Vasisht:2011p6552} and SiO$_2$.\cite{Garofalini:1990p4811}

\begin{figure}
\begin{center}
\includegraphics[ ]{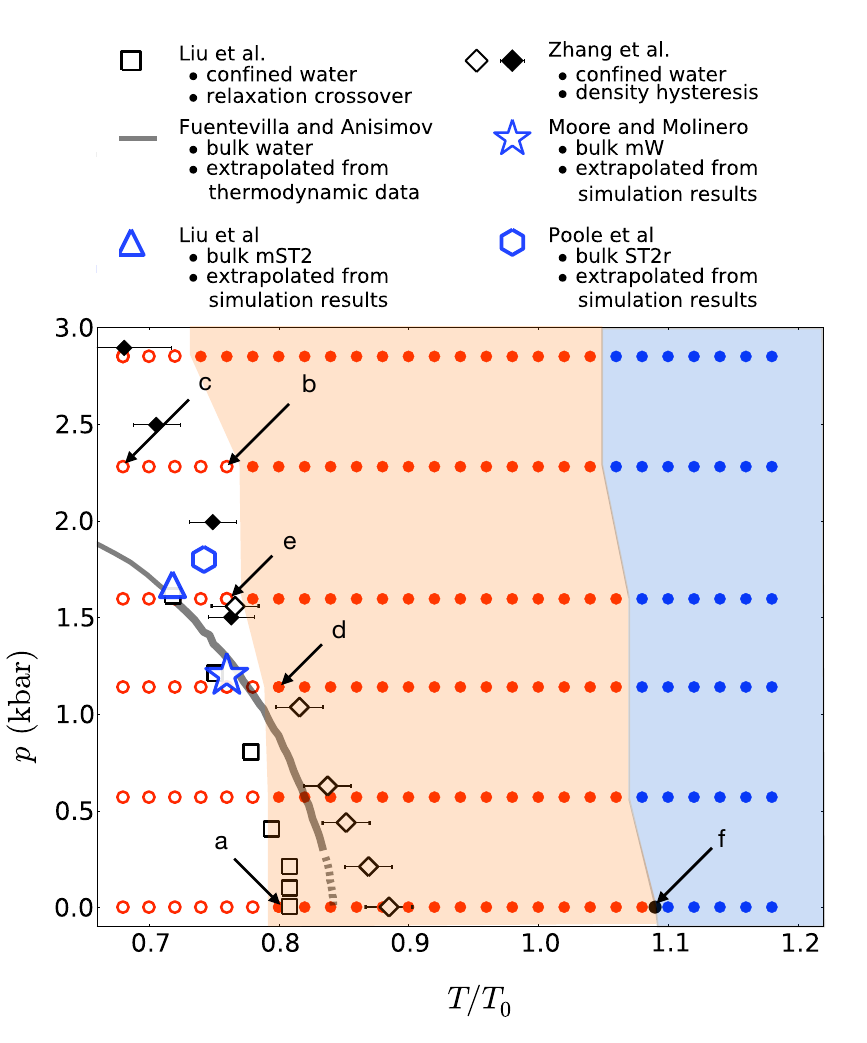}
\caption{Phase space sampled in our calculations for the mW model. Open circles refer to states where the amorphous liquid is found to be unstable, filled red circles refer to states where the amorphous liquid is found to be metastable with respect to the crystal, and filled blue circles refer to states where the liquid is found to be stable with respect to the crystal. The labeled circles (a), (b) and (c) identify state points where explicit free energy surfaces are shown in Fig. ~\ref{Fi:fe_q6rho}. The circles labeled (d) and (e) identify state points where explicit free energy surfaces are shown in Fig. ~\ref{Fi:fe_psi6rho}. The black circle labeled (f) is the state point where crystal-liquid phase coexistence  is examined in Fig.  ~\ref{Fi:scaling}.  Lines, diamonds and squares locate previous estimates of a liquid-liquid phase transition inferred from experimental results;\cite{Liu:2005p2412,Zhang:2010p4637,Fuentevilla:2006p4635} see text.  A star locates a previous prediction of a liquid-liquid critical point based upon extrapolation of simulation results for the mW model.\cite{Moore:2009p248} The blue triangle and hexagon are estimates of low temperature critical point locations obtained from interpreting simulation results for the mST2~\cite{Liu:2009p776} model and the ST2r~\cite{Poole:2005p6308} model, respectively.}  \label{Fi:phase_d}
\end{center} 
\end{figure}

\section{Phase space studied}

The mW model \cite{Molinero:2008p4576} differs from the Stillinger-Weber inter-particle potential energy function for silicon in two ways.  The first is the adoption of different values for the length and energy parameters of the model.  This difference is inconsequential to our study because it amounts to a simple rescaling of temperature and density.  The second is more substantive but slight.  Specifically, to capture some thermodynamic properties of water, the mW model has a partitioning between two- and three-body terms that differs by 10\% from the partitioning for silicon.  See Ref. \onlinecite{Molinero:2008p4576} for details concerning the mW model.

\subsection{Pressure-temperature phase diagram of the mW model}

We have studied the phase behavior of the mW model by computing free energy surfaces throughout its condensed phases.  Figure \ref{Fi:phase_d} shows the state points examined. Each circle represents a state point where the free energy has been calculated as a function of the global system density and an order parameter that quantifies broken orientational symmetry. The latter distinguishes an amorphous liquid phase from a crystal, whereas the former distinguishes two amorphous phases with different densities. Basins in the free energy surface establish relative stabilities of the phases.  

To put this diagram in context, we highlight state points that others have identified as relevant to a liquid-liquid phase transition in supercooled water.  The temperature of maximum density at low pressure sets the scale of the figure.  This chosen reference temperature is $T_0 = 250$K  for the mW model,\cite{Molinero:2008p4576} and it is $T_0 = 277$K for water.\cite{NBS_steam} The phase diagram in Fig. \ref{Fi:phase_d} shows that the density maximum of liquid mW occurs at slightly supercooled conditions while that of experimental water occurs at a temperatures slightly higher than the freezing temperature. 

The points identified by Liu et al.~\cite{Liu:2005p2412} come from measured relaxation times of water confined in silica nanopores with a 7\AA\,  radius.  These relaxation times have a temperature dependence that changes from super-Arrhenius to Arrhenius upon cooling below a crossover temperature, $T_\mathrm{x}(p)$.  This temperature depends upon external pressure $p$, and points on this line are shown in Fig.~\ref{Fi:phase_d} with unfilled squares, which are attributed in Ref.~ \onlinecite{Liu:2005p2412} to crossing a ``Widom line". A Widom line refers to a locus of maximum response that ends at a critical point.\cite{Kumar:2005p2095}  In cases where a phase transition exists, there are many such lines because different response functions have different lines of extrema.  Ambiguity ceases only in the proximity of a critical point.  But whether any such lines can be related to $T_\mathrm{x}(p)$ is unclear because Widom lines refer to time-independent thermodynamic behavior and $T_\mathrm{x}(p)$ refers to time-dependent non-equilibrium behavior.

A different basis for identifying relevant points is made by Zhang et al. considering the same system.\cite{Zhang:2010p4637}  In this case it is the density of the water that is measured. This observed density exhibits hysteresis upon alternating heating and cooling scans, and the hysteresis grows upon increasing pressure. We have already noted that it is questionable whether the phase behavior of bulk water can be related to that of water confined to narrow pores.  Virtually all molecules in those pores are influenced by interfaces. Nevertheless, points of maximum hysteresis, denoted by the filled diamonds in Fig.~\ref{Fi:phase_d}, have been attributed to a line of first-order liquid-liquid transitions, and points of less significant hysteresis, marked by open diamonds in the figure, are attributed to a continuation of that transition.\cite{Zhang:2010p4637}

Another proposed line of liquid-liquid transitions is constructed by Fuentevilla and Anisimov.\cite{Fuentevilla:2006p4635}  Here, a postulated scaling form is used to extrapolate from experimentally accessible equilibrium thermodynamic data.  The resulting prediction and its analytic continuation are drawn as solid and dashed lines in Fig.~\ref{Fi:phase_d}. Even if a critical point is present this predicted line is questionable because it is generally impossible to identify critical divergences from a small rise in noncritical background fluctuations of the sort contributing to the heat capacity at standard conditions. This fact is illustrated by Moore and Molinero's predicted critical point for mW water.\cite{Moore:2009p248}  Its location, the star in Fig.~\ref{Fi:phase_d}, is found by extrapolation from a small rise in a response function computed at distant thermodynamic conditions.  We find no evidence for a liquid-liquid transition anywhere near this predicted critical point.  Rather, it and all other estimates pertaining to a purported liquid-liquid transition lie close to a spinodal associated with crystalization.  This finding is not inconsistent with Moore and Molinero's more recent report that the amorphous phase of the mW model seems to be forever changing and impossible to equilibrate at a point in the phase diagram where liquid-liquid transitions have been suggested.~\cite{Moore:2010p1923}

Two additional marked points in Fig.~\ref{Fi:phase_d}, the blue triangle and hexagon, refer to other estimates of a location for a liquid-liquid critical point.  These are estimates obtained from extrapolating simulation results for variants of the ST2 water model,\cite{Stillinger:2003p2558} about which we have more to say later.

\begin{figure*}[ht]
\begin{center}
\includegraphics[scale=0.35]{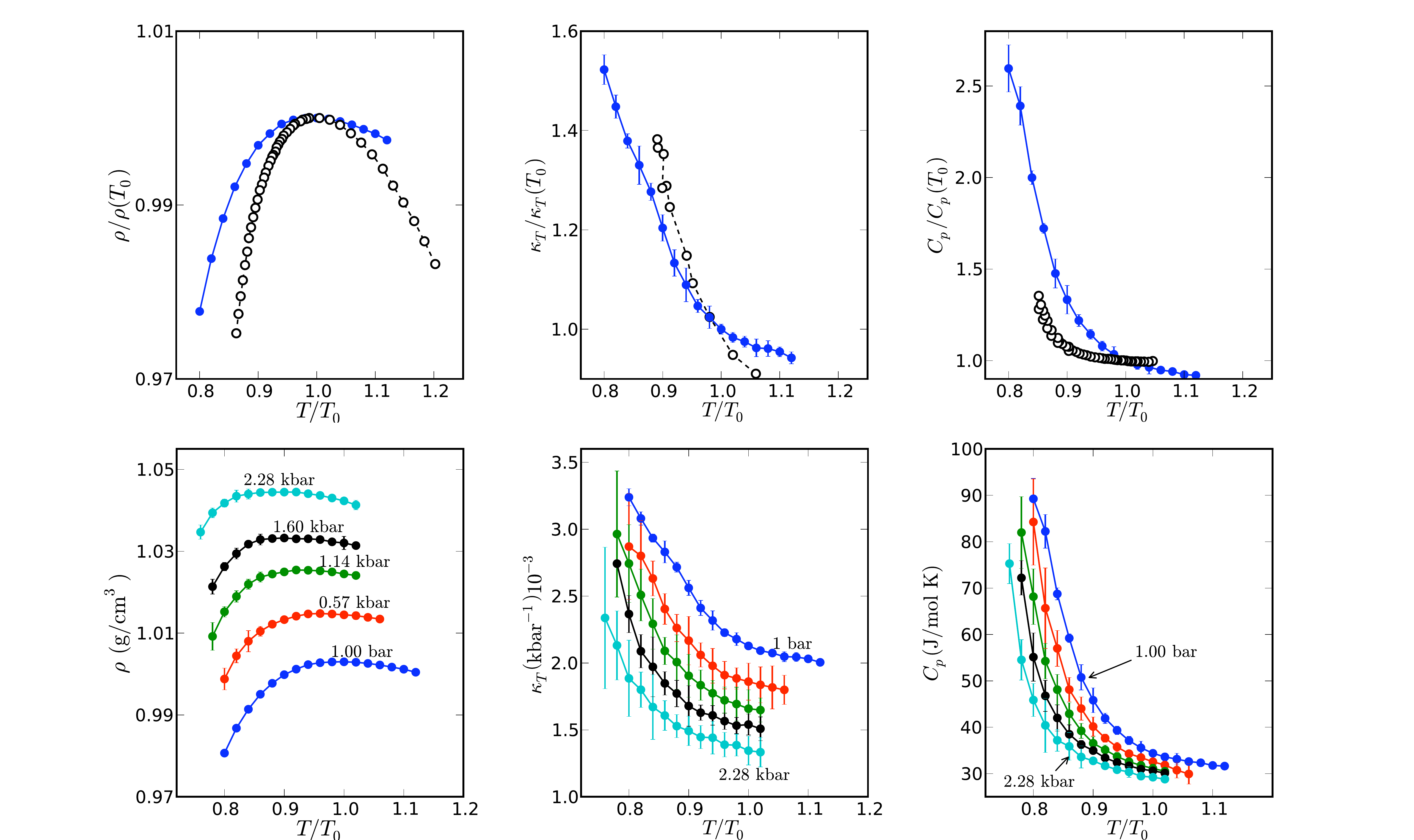}
\caption{Average thermodynamic properties as a function of temperature.  Upper panels compare mW model results with those of experiment at $p=$ 1 bar. The filled blue circles are the calculated results for the mW model, where error estimates are one standard deviation. The empty black circles are experimental results taken from Ref.~\onlinecite{Debenedetti:2003p813}.  Lower panels show pressure dependence of the mW model results.}\label{Fi:state_prop}
\end{center} 
\end{figure*}

\subsection{Anomalous thermodynamics of the mW model}

Water exhibits anomalous thermodynamic properties at low temperatures, properties that are are non-singular but nonetheless unusual. Because these behaviors have been proposed as indicators of a liquid-liquid transition, it is important to show that the mW model exhibits such behaviors. Specifically, we focus on the density maximum as a function of temperature, and the relatively large rate of increases upon lowering temperature of both isothermal compressibility and isobaric heat capacity.\cite{Debenedetti:2003p813} 
 
Thus, we have used a constant pressure ensemble to compute $\rho = N/\langle V \rangle$, $ \kappa_T =  \langle \left (\delta V \right )^2 \rangle / k_\mathrm{B} T \langle V \rangle $ and $C_p = \langle \left (\delta H \right )^2 \rangle / k_\mathrm{B} T^2$ for the mW model.  Here, $N$, $V$ and $H$ denote number of molecules, volume and enthalpy, respectively; $\delta V$ and $\delta H$ denote deviations from mean values of $V$ and $H$, respectively; the pointed brackets denote an ensemble average; $k_\mathrm{B}$ is Boltzmann's constant. 

Figure~\ref{Fi:state_prop} compares our computed results at ambient pressure with those found from experimental observation of water.\cite{Debenedetti:2003p813}  As in Fig.~\ref{Fi:phase_d}, we use the low pressure point of density maximum as our reference state for these comparisons. Figure~\ref{Fi:state_prop} shows that the qualitative trends and magnitude of anomalies of mW water agree with those of experimental water.  The low-temperature end of the displayed graphs occur at the point where the liquid becomes unstable.  Down to that temperature, the growths of $\kappa_T$ and $C_p$ are notable but modest in size and far from the sort of divergent behavior one would ordinarily associate with a critical point or phase boundary.  At all stable and meta-stable liquid phase states we have studied, see Fig.~\ref{Fi:phase_d}, we find similar nonsingular behavior.

The lower panels of Fig.~\ref{Fi:state_prop} show that the trends observed at 1 bar persist to higher pressures in the mW model, with the density maximum temperature decreasing slightly as pressure increases.  These trends are consistent with experiment.\cite{Debenedetti:2003p813} 

\section{Order parameters and free energies}
In this section, we define order parameters and present free energy functions of those order parameters.

\subsection{Measures of crystalline order}
We use two types of order parameters.  One is bulk density, the other quantifies orientational order. For the latter, we use Steinhardt, Nelson and Ronchetti's $Q_6$ and $\psi_6$.\cite{Steinhardt:1983p1432}  For a finite system analyzed with computer simulation, these variables prove more convenient than Fourier the components of the density.  
They also prove more useful than dynamic measures, which cannot distinguish liquid from crystal at supercooled conditions, where diffusion is slow in the liquid due to glassy dynamics and nonzero in the crystal due to defect motion.

Both $Q_6$ and $\psi_6$ are functions of a projection of the density field into averaged spherical harmonic components.  To evaluate $Q_\ell$, for each water molecule $i$, we calculate the set of quantities
\begin{equation}
q_{\ell,m}^i= \frac{1}{4} \sum_{j\in n_i}^4 Y_\ell^m (\phi_{ij},\theta_{ij}) \,,\quad -\ell \leqslant m\leqslant \ell\, ,
\end{equation}
where the sum is over those nearest 4 neighbors, $n_i$. $Y_\ell^m (\phi_{ij},\theta_{ij})$ is the $\ell,m$ spherical harmonic function associated with of the angular coordinates of the vector  $\vec{r}_{i} - \vec{r}_j $ joining molecules $i$ and $j$, measured with respect to an arbitrary external frame. Since $q_{\ell ,m}^i$ is defined in terms of spherical harmonics, it transforms simply under rotations of the system or the arbitrary external frame.  These quantities are then summed over all particles to obtain a global metric  
\begin{equation}
Q_{\ell,m}= \sum_{i=1}^N q_{\ell,m}^i \,,
\end{equation}
and then contracted along the $m$ axis to produce a parameter that is invariant with respect to the orientation of the arbitrary external frame,
\begin{equation}
Q_\ell = \frac{1}{N} \left(  \sum_{m=-\ell}^\ell Q_{\ell,m}Q_{\ell,m}^* \right)^{1/2} \,.
\end{equation}

The other orientation order parameter we consider, $\psi_\ell$, is evaluated by first defining bond variables through local contractions of the $q_{\ell,m}$, which are reference frame independent, 
\begin{equation}
b_{ij}= \frac{\sum_{m=-\ell}^\ell q_{\ell,m}^i q_{\ell,m}^{j*}}{\left( \sum_{m=-\ell}^\ell q_{\ell,m}^i q_{\ell,m}^{i*} \right)^{1/2} \left( \sum_{m=-\ell}^\ell q_{\ell,m}^j q_{\ell,m}^{j*} \right)^{1/2}} ,
\end{equation}
and then summing over all of the bonds made between molecule $i$ and its nearest 4 neighbors,
\begin{equation}
\psi_\ell^i = \frac{1}{4}  \sum_{j\in n_i}^4 b_{ij}
\end{equation}
Finally, the global parameter is obtained by summing over all molecules,
\begin{equation}
\psi_\ell= \frac{1}{N}  \sum_{i=1}^N \psi_\ell^i
\end{equation}

\begin{figure*}[ht]
\begin{center}
\includegraphics[]{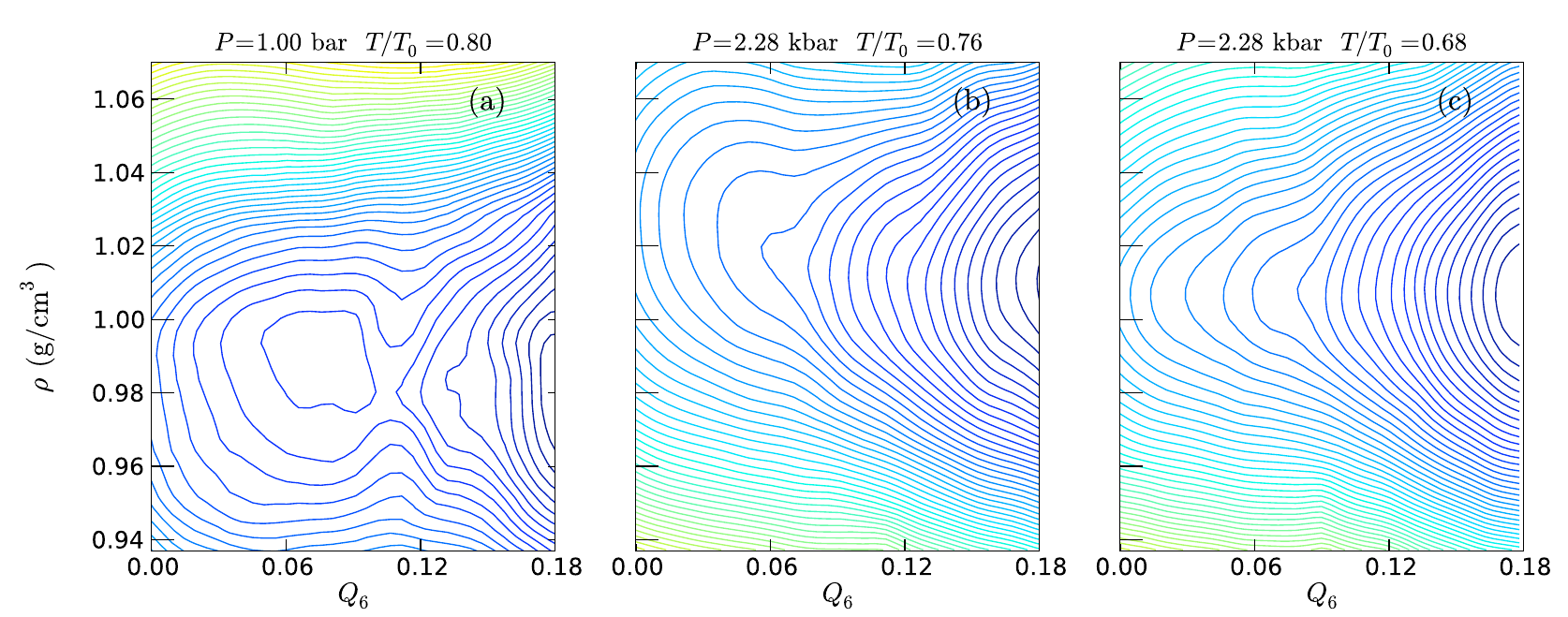}
\caption{Free energy surfaces for mW water as a function of $\rho$ and $Q_6$. The system is periodically replicated and contains $N=216$ particles. As shown in (a),  the liquid is metastable with respect to the crystal. As the system is cooled, the barrier disappears, as illustrated in (b). Finally in (c) the free energy obtains a large gradient along the $Q_6$ direction and fluctuations in density are damped out. Adjacent contour lines are spaced by 1 $k_\mathrm{B} T$, and statistical uncertainties are smaller than that energy.}~\label{Fi:fe_q6rho}
\end{center} 
\end{figure*}

The mean or most probable value of $Q_{\ell}$ for an amorphous phase approaches zero in the thermodynamic limit, while it is finite for a crystalline phase.  As such, $Q_\ell$ is a distinguishing order parameter for amorphous and crystalline phases.  In contrast, because its contractions occur locally and not over the entire system, $\psi_{\ell}$ is non-vanishing in the thermodynamic limit for both disordered and ordered states.  Nevertheless it is a useful measure of orientational order because the distributions of $\psi_\ell^i$ for the low temperature liquid are sensitive to the amount of crystallization in the system, and their mean values at low temperatures differ significantly between liquid and crystal.  Further, as  $\psi_{\ell}$ retains local information, it is useful in determining the existence of grain boundaries and defects.  We have taken the $\ell$ = 6 multipole because we have found empirically that it is particularly sensitive to distinguishing liquid water and ice.  

\subsection{Free energy surfaces at conditions of metastability}

Free energies of density $\rho$, orientational order parameters $Q_6$ and $\psi_6$, and so forth, are related to the probabilities of the order parameters in the usual way. Specifically, 
\begin{equation}
\label{Eq:fe}
F(\rho,Q_6, ...) =- k_\mathrm{B} T \ln  P \left (\rho,Q_6, ... \right ) + \, \mathrm{const.}
\end{equation}
where the probability $ P \left (\rho,Q_6, ... \right ) $ is proportional to the partition function for micro-states with the specified values of the order parameters. The irrelevant additive constant in Eq.~\ref{Eq:fe} refers to normalization and standard state conventions. 

To evaluate the probabilities and their associated free energies, we have adopted a hybrid Monte Carlo simulation approach as used by Duane et. al.\cite{Duane:1987p3223} We consider ensembles with $N$, $p$, and $T$ fixed. Two different moves are made within this framework: random changes in volume, and short molecular dynamics trajectories. These are made with a ratio of 1:5. Maximum volume displacement  and maximum molecular dynamics trajectory length are adjusted to yield a 30\% acceptance. This technique produces suitably swift equilibration even within the supercooled regime. 

The order parameters $\rho$, $Q_6$ or $\psi_6$, are controlled with umbrella sampling, by propagating the system under its unbiased hamiltonian and computing the order parameters only when determining Metropolis acceptance probabilities. All molecular dynamics propagation was done using the LAMMPS molecular dynamics simulation package.\cite{Plimpton:1995p3851} Most of the free energy calculations were accomplished with 216 particles. For each window in the umbrella sampling, the simulations ran long enough to obtain at least 1000 independent samples of each of the biased observables. The umbrella biasing potentials employed were 
\begin{equation}
\Delta U= k (\rho-\rho^*)^2 + \kappa (Q_6-Q_6^*)^2 
\end{equation}
or  the same formula with $Q_6$ replaced by $\psi_6$. Adopting $\kappa$ in the range of  500 to 2000 $k_\mathrm{B} T $ and  $k$ in the range of 1000 to 2000 $k_\mathrm{B} T\,  \mathrm{cm}^3 / \mathrm{g} $ proved satisfactory. Statistics gathered in these biased ensembles were unweighted and the free energy differences between each ensemble were estimated using multi-state Bennett acceptance ratio (MBAR).\cite{Shirts:2008p4360} Error estimates for the free energies we have calculated in this way are less than $k_\mathrm{B} T$.
 
 \begin{figure}[b]
\begin{center}
\includegraphics[ ]{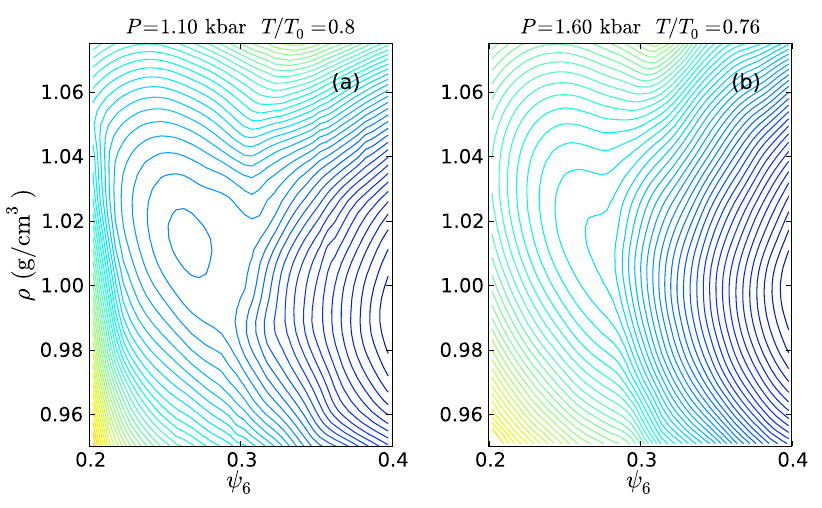}
\caption{Free energy for mW wateras a function of $\rho$ vs $\psi_6$ at conditions where the liquid is (a) metastable, and (b) unstable. The system is periodically replicated and contains $N=216$ particles. Adjacent contour lines are spaced by 1 $k_\mathrm{B} T$, and statistical uncertainties are smaller than that energy.}
\label{Fi:fe_psi6rho}
\end{center}
\end{figure}

Figure ~\ref{Fi:fe_q6rho} depicts representative free energies for three different state points. The locations of those state points are noted in Fig.~\ref{Fi:phase_d}. Each free energy surface includes the range of densities where liquid and crystal basins are located. With the variable $Q_6$, we see a significant separation between liquid and crystal basins. For the state points considered in Fig.~\ref{Fi:fe_q6rho}, with $N=$ 216, the crystal basin is centered around $Q_6 \approx$ 0.5, while the liquid basin, when it exists, is centered around  $Q_6 \approx$ 0.05. As $N$ increases, the former changes little, but the latter tends to zero. This behavior is illustrated explicitly in the next section.

The state points considered in Fig.~\ref{Fi:fe_q6rho} show how the free energy surfaces evolve as the pressure or temperature are changed. In Fig.~\ref{Fi:fe_q6rho}(a), a barrier separates the liquid phase from the crystal. Therefore, at that state point the liquid is metastable. Lowering the temperature and increasing the pressure, Fig.~\ref{Fi:fe_q6rho} (b) shows the barrier to crystallization has vanished. Further decreasing temperature, Fig.~\ref{Fi:fe_q6rho} (c) shows increasing driving force towards the stable crystal. At those state points, the liquid is unstable.

Similar behavior is found with the free energy of $\rho$ and $\psi_6$. This function, $F(\rho,\psi_6)$, is shown in Fig.~\ref{Fi:fe_psi6rho} at two different state points.  Figure \ref{Fi:fe_psi6rho}(a) shows this free energy at a temperature and pressure, labeled (d) in Fig. ~\ref{Fi:phase_d}, where the liquid is metastable with respect to the crystal.  At this state point, the mean value, $\langle \psi_6 \rangle$ is about 0.27, a value that reflects the relatively small amount of local ordering present in the supercooled liquid. In contrast, for the crystal we find $\langle \psi_6 \rangle \approx 0.9$. Fig. \ref{Fi:fe_psi6rho}(b) shows the free energy for a temperature and pressure in the region of the phase diagram where the amorphous phase is unstable, the so-called no man's land. This point, labeled (e) in Fig.~\ref{Fi:phase_d}, is close to a proposed location of a liquid-liquid critical point.\cite{Moore:2009p248} We see, however, that it is not a point of criticality. The behavior of $\psi_6$ is strongly correlated to the potential energy. This fact follows from the functional form of $\psi_6$ and the three-body potential of the mW model. Thus, the behavior of $F(\rho,\psi_6)$ should be similar to that of $F(\rho,U)$ where $U$ denotes the total potential energy of the mW model. 

For all of the state points considered, which includes a broad swath of  no-man's land, there is no evidence of a bifurcation of the free energy along the density direction within the liquid region (i.e., where $Q_6$ and $\psi_6$ are small). What bifurcation does exist is associated with a transition between an amorphous phase and a crystal. We now consider whether it is a first order transition.

\section{Freezing transition}
\subsection{mW Model}
 
\begin{figure}
\begin{center}
\includegraphics[ ]{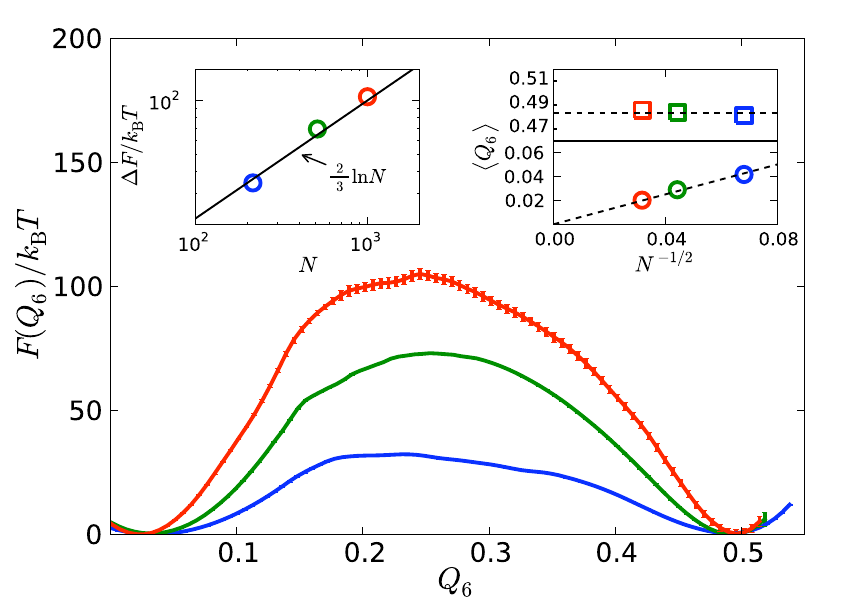}
\caption{Free energies as a function of $Q_6$ for $N=$ 216 (blue), 512 (green), and 1000 (red), calculated at $T/T_0$=1.09 and $p=$1 bar. Left inset: Interfacial free energy for different system sizes. For comparison a line of slope $2/3$ is also shown. Right inset: The mean value of $Q_6$ for liquid (circles) and crystal (squares) for different system sizes. Error estimates are shown in the main figure, but are smaller than the symbols in the insets.}\label{Fi:scaling}
\end{center} 
\end{figure}

 The character of the transition can be analyzed by studying the system-size dependence of the contracted free energy.\cite{Binder_book}
 \begin{equation}
F (Q_6) = -k_\mathrm{B} T \ln \left ( \int \mathrm{d} \rho \,  \exp{\left [-\beta F(\rho,Q_6) \right ]} \right ).
\end{equation}
This function is shown in Fig. \ref{Fi:scaling} for the mW model at one of the pressures and temperatures where an amorphous phase is in coexistence with the crystal. Such points are at the boundary between the blue and red regions in Fig. \ref{Fi:phase_d}. The quantity $\Delta F(Q_6) = F(Q_6) - \mathrm{min} [ F(Q_6)]$ reaches its maximum value when an interface separating amorphous and crystal phases extends across the entire system. This maximum value is the interfacial free energy. Accordingly, for a first-order transition, it should be proportional to $N^{2/3}$. This scaling is satisfied to a good approximation for the system sizes considered in Fig. \ref{Fi:scaling}.  

Nonzero values of $Q_6$ in an amorphous phase are due to fluctuations. As such, the mean value of $Q_6$ for the amorphous phase should disappear as $1/N^{1/2}$. This scaling is also found for the system sizes studied and is illustrated in  Fig. \ref{Fi:scaling}. In contrast, for a crystal $Q_6$ will have a nonzero mean that remains finite as $N \rightarrow \infty$. This behavior is consistent with our numerical results, as also illustrated in Fig. \ref{Fi:scaling}.

Thus, the transition between liquid and crystal in mW water appears to be a standard freezing transition which is first order and between phases with different orientational symmetry.  The analysis used here to reach that conclusion can be applied to other models.  For example, we have carried out this analysis to study the Stillinger-Weber model of silicon.  Here too, we find that the model exhibits a freezing transition, and contrary to recent suggestions\cite{Vasisht:2011p6552} there is no evidence for an equilibrium liquid-liquid transition. We also arrive at this same conclusion for another model of water, which we turn to now.

\subsection{mST2 model}

\begin{figure*}
\begin{center}
\includegraphics[ ]{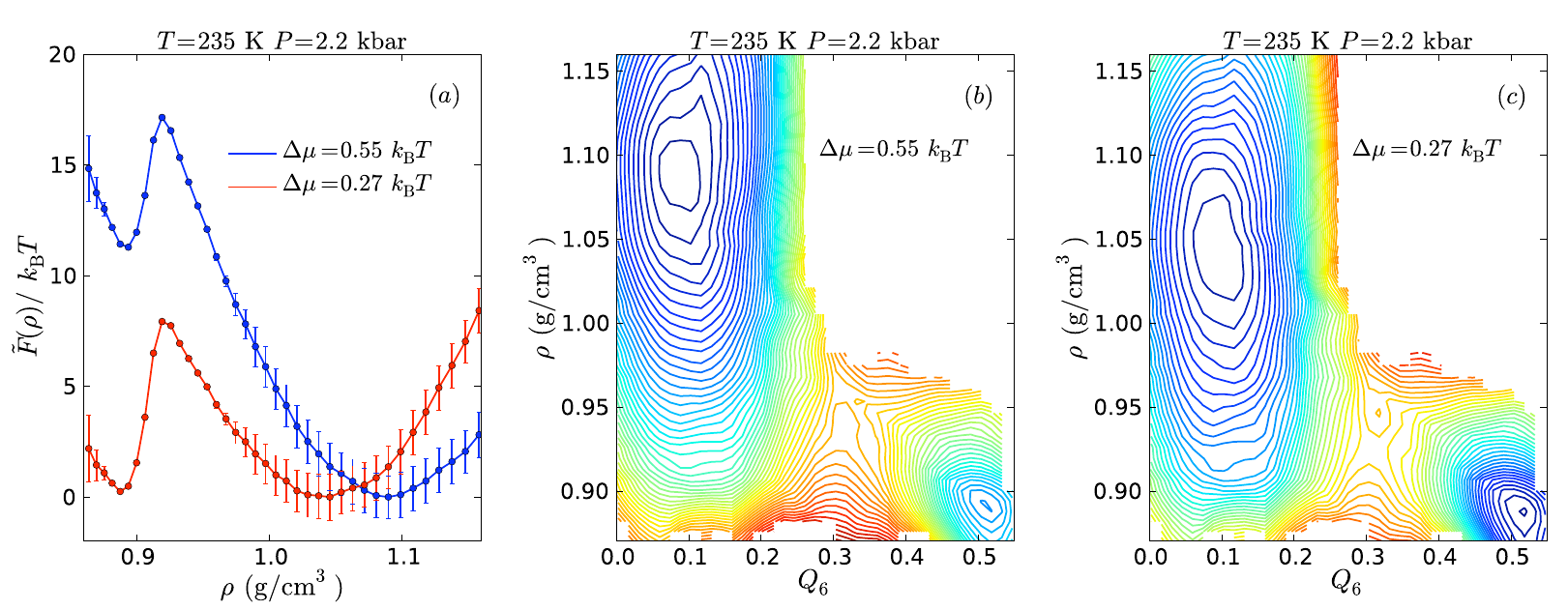}
\caption{Free energies for the mST2 model of water. The system is periodically replicated and contains $N=216$ molecules. Panel (a) is the contracted $\tilde{F}(\rho)$. Panels (b) and (c) are the surfaces $\tilde{F}(\rho,Q_6)$. Phase coexistence between amorphous and crystal phases occurs at $\Delta \mu=0.27 k_\mathrm{B} T$, where $\Delta \mu$ is the chemical potential relative to that of phase space point $(T,p)$=(235 K,2.2 kbar). Adjacent contour lines in (b) and (c) are spaced by $1\, k_\mathrm{B}T$ and statistical uncertainties are of the order of, or less than, that energy. Error bars in (a) are one standard deviation.\label{Fi:st2}}
\end{center} 
\end{figure*}

We have considered molecules interacting by a modified form of Stillinger and Rahman's pair potential.\cite{Stillinger:2003p2558} The modification incorporates long-ranged electrostatics rather than the simple spherical truncation of the original model. The resulting system, which we call ``mST2 water," has been studied by Liu et al.,\cite{Liu:2009p776} who report finding evidence of a liquid-liquid transition and an associated critical point. The mST2 model is more difficult to simulate than the mW model because the former contains long-ranged interactions and the latter does not. As such, our investigation of its behavior is more limited than those we have preformed for mW water. Nevertheless, our investigation seems sufficient to challenge the finding of a liquid-liquid transition at the conditions examined by Liu et al. It also seems sufficient to discount an assortment of less direct simulation studies that also report evidence of a liquid-liquid transition in ST2 water.\cite{Kumar:2005p2095,Harrington:1997p202,Harrington:1997p202,Poole:2005p6308,Poole:1993p5450}  Indeed, we have considered this particular model of water because it is so often examined in publications supporting the hypothesized liquid-liquid transition, most recently in a paper\cite{PooleArXiv} motivated by preliminary reports of our work.

Figure \ref{Fi:st2} shows free energies we have computed for the mST2 model using $N=216$ molecules. The procedures we employed are identical to those used for the mW model, except for the technical detail that we modify LAMMPS to handle the specific mST2 potential. We focus on the region of the $p$-$T$ plane where Liu et al. report bifurcation in the free energy as a function of density. In that region, we too find a bifurcation, but not between two amorphous phases. The grand canonical Monte Carlo simulation method of Ref.~ \onlinecite{Liu:2009p776} is sufficient to detect a phase boundary for the liquid, but it cannot distinguish liquid from crystal because it does not control distinguishing order parameters. In our calculations, where both $\rho$ and $Q_6$ are controlled, we find that a boundary does in fact exist between liquid and crystal. But at the thermodynamic conditions considered by Liu et al., there is no evidence of a second liquid basin in the free energy $F(\rho,Q_6)$. 

The specific free energies shown are found by first computing $F(\rho,Q_6)$ from our simulations at $T=235$ K and $p=$ 2.2 kbar, i.e. we compute $F(\rho,Q_6)=F(\rho,Q_6;p,T)$. The free energy shows that for this point in the phase diagram, the crystal is stable with respect to the liquid. A specific state point considered by Liu et al. is at the same temperature but a different pressure or chemical potential for which the free energy can be reached by a shift in chemical potential
\begin{equation}
\tilde{F}(\rho,Q_6;T,\Delta \mu) = F(\rho,Q_6) - \rho \bar{V} \Delta \mu ,
\end{equation}
with $\Delta \mu \approx \,  0.55 \, k_\mathrm{B}T$. Here, $\Delta \mu$ is the chemical potential relative to that at $(T,p)$=(235 K, 2.2 kbar), and $\bar{V}$ is the average volume of the system at that temperature and pressure. A lower value of $\Delta \mu$ brings the system to a point of coexistence between the liquid and the crystal. These free energy surfaces are shown in Panels (b) and (c) of Fig. \ref{Fi:st2}. The free energy computed by Liu et al. is the contraction 
\begin{equation}
 \tilde{F} (\rho) = - k_\mathrm{B} T  \ln \left ( \int \mathrm{d}Q_6 \,  \exp{\left [-\beta F(\rho,Q_6) - \beta  \rho \bar{V} \Delta \mu \right ]} \right ) 
\end{equation}
This function is shown in Panel (a) of Fig.~\ref{Fi:st2}. 

Like Liu et al, we find a bistable free energy at this temperature and for this size system.  The locations for the minima we find for $\tilde{F}(\rho)$ are in good accord with those found by Liu et al. But our free energy has a large barrier between the two basins, reflecting a finite crystal-liquid surface tension, while that reported by Liu et al exhibits a small barrier.  Liu et al. suggest that their result is indicative of a liquid-liquid transition and the proximity of a critical point.  However, our free energy surface shows no such phase transition behavior.  There is only a crystal-liquid first-order transition.  We suggest that the Liu et al. result is a non-equilibrium phenomenon, where a long molecular dynamics run at constant $T-p$ and initiated from their low-density amorphous phase will eventually equilibrate in either the low density crystal or in the higher density metastable liquid.  The time scale for this equilibration is long, as we discuss in the next section.  

Whatever the cause for the Liu et al results, the bi-stability cannot be attributed to a liquid-liquid transition without also showing that the barrier separating presumed liquid-phase basins satisfies the requisite growth with $N$, scaling as $N^{2/3}$.  This demonstration has not been done, and from our results, it is unlikely that it can be done.

Another variant of the ST2 model, considered by Poole et al.,\cite{Poole:2005p6308} uses a reaction field approximation to estimate the effects of long-ranged forces.  We call it the``ST2r'' model.  One expects similar phase behaviors from the mST2 and ST2r models.\cite{PoolePrivate} Based on an extrapolation from the equation of state computed for ST2r model, Poole et al. predict the presence of a liquid-liquid transition, and the critical point location obtained from that estimate is shown in Fig.~\ref{Fi:phase_d}.  The density-maximum reference temperature for both mST2 and ST2r is $T_0 \approx 330 \,\mathrm{K}$.   Poole et al.'s estimate the critical temperature to be $T_\mathrm{c}=245\,$K.  Our calculations for mST2, shown in Fig.~\ref{Fi:st2}, are at the lower temperature, $T=235\,$K.  Accordingly, at some pressure, we should find bistable liquid behavior if indeed a critical point existed at the higher temperature.  But we find that upon adding $\Delta p \,V$ to our computed $F(\rho, Q_6; 2.2\, \mathrm{kbar}, 235 \,\mathrm{K})$, where $\Delta p = p - 2.2 \,\mathrm{kbar}$, no second liquid basin can be discerned for any reasonable value of $p$. Therefore, and similar the to behavior found with the mW model, extrapolation from the behavior of a one-phase system as done in Ref.~\onlinecite{Poole:2005p6308} proves to be a poor indicator of a phase transition. 

\section{Dynamic Metastability}

To arrive at the results of the previous sections, equilibration is achieved with umbrella sampling.  Various other reweighting Monte Carlo procedures could be used.\cite{Binder_book, Frenkel_book}   Some researchers, however, attempt to learn about a possible reversible phase transition in supercooled water through straightforward molecular dynamics simulation.  This approach is limited to cases where relaxation is swift compared to computationally feasible trajectory lengths, but relaxation associated with phase transitions is generally not swift, especially at supercooled conditions.  To judge the feasibility of such an approach, it is therefore useful to estimate pertinent relaxation times.  For supercooled water, there are two important classes:  times required to nucleate and grow a crystal, and times required to reorganize atomic arrangements in the liquid.  We have estimated both with molecular dynamics of mW water.  We use the equilibrium sampling described in prior sections to prepare initial configurations from which we carry out Newtonian trajectories to compute dynamical properties.  These trajectories evolve with a Nose-Hoover integrator\cite{Martyna:1994p4409} with a thermostat time constant of $5\,$ps  and a barostat time constant of $5\,$ps.

\begin{figure}[ht]
\begin{center}
\includegraphics[ ]{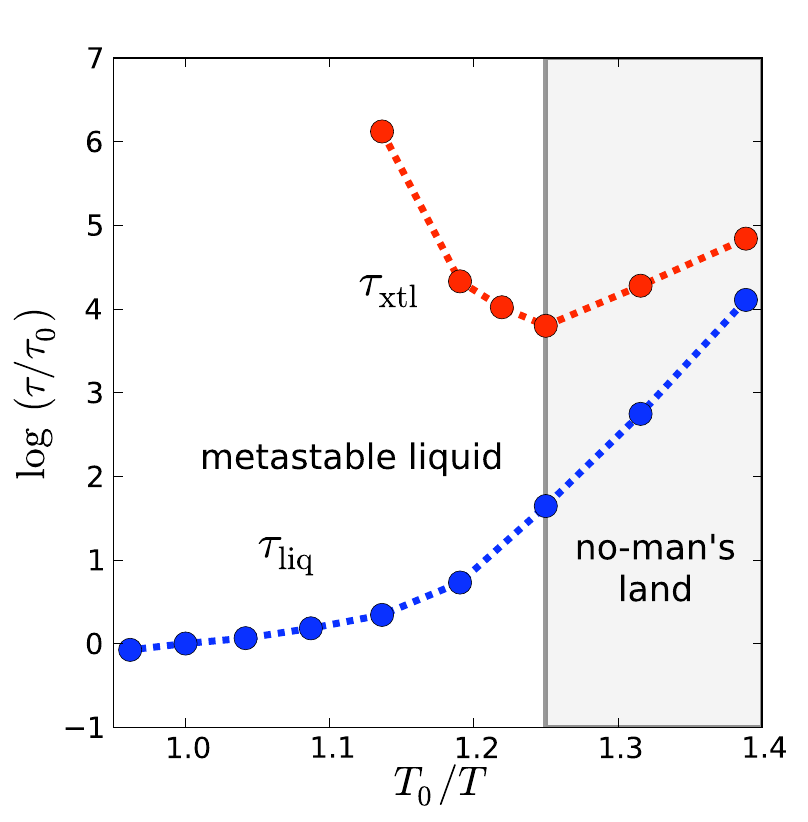}
\caption{Time scales of the supercooled liquid mW water at 1 bar. Computed structural relaxation times, $\tau_\mathrm{liq}$, are shown with blue points. Computed crystallization times, $\tau_\mathrm{xtl}$, are shown with red points. Statistical uncertainties are smaller than the symbols. Dashed lines are drawn as guides to the eye. The grey region is where the liquid is unstable.
}\label{Fi:tau_temp}
\end{center} 
\end{figure}

At conditions of liquid metastability, where a free energy barrier separates liquid and crystal basins, nucleation is the rate-determining step to form the equilibrium phase.  For those conditions, we have computed this rate constant following a standard Bennett-Chandler procedure for rare-event sampling.\cite{Frenkel_book}  Specifically, we take $Q_6$ as the reaction coordinate, so that the rate constant for nucleation is $k_\mathrm{nuc} = \nu \exp[-F(Q_6^*)/k_\mathrm{B}T]$, where $Q_6^*$ is the point of maximum $F(Q_6)$ between liquid and crystal basins, and the prefactor, $\nu$, includes the transmission coefficient.  This prefactor is determined by sampling trajectories initialized at the top of the free energy barrier, i.e., initialized at configurations with $Q_6 = Q_6^*$.\cite{TenWolde:1995p4499}  Other choices of transition state are possible, but the net result is invariant to that choice.\cite{Chandler:1978p6240}  The mean time to nucleate the crystal is then $1/k_\mathrm{nuc}= \tau_\mathrm{xtl}$.  Results obtained in that way with $N=216$ mW particles are shown in Fig. \ref{Fi:tau_temp}.  The reference time used to represent these results, $\tau_0$, is the structural relaxation time at the reference liquid state used throughout this paper, $T_0=250 K$.  For mW this time is $\tau_0 \approx 0.5$ ps; for experimental water, this time is larger by a factor of about 5. 

For conditions of liquid instability, (i.e. the no-man's land where there is no barrier between liquid and crystal), the method of rare-event sampling is no longer appropriate.  For those conditions, we compute first-passage times.\cite{Van_Kampen_book}  The results obtained depend upon the initial preparation of the system because the unstable system is far from equilibrium.  In the particular preparation we employ, we equilibrate the system in the liquid region at $T/T_0=0.84$  where the liquid is metastable. Then at time $t=0$, the system is quenched to the target temperature and allowed to evolve towards the crystal state. The first-passage time is taken as the first time a trajectory with initial conditions prepared in that way reaches a configuration with $Q_6 = 0.2$.  We find an exponential distribution of first-passage times. Mean values from that distribution are the values of $\tau_\mathrm{xtl}$ shown in Fig. \ref{Fi:tau_temp} for no-man's land state points.  

The line in the $p$-$T$ plane separating filled and unfilled circles in Fig. 1 is the boundary between metastable and unstable liquid conditions. At metastable conditions close to that boundary, we have checked that the $\tau_\mathrm{xtl}$ found from the first-passage method agrees with that found from the rare-event sampling. 

For the structural relaxation time of the supercooled metastable liquid, we consider trajectories initiated from equilibrated configurations in the liquid region and observe the time $t$ it takes an average particle to move one molecular diameter, i.e., 3\AA =$(1/N)\sum_i |\vec{r}_i(t)-\vec{r}_i(0)|$. For temperatures below the limit of liquid stability, we use the same procedure for generating initial conditions as we used in the calculation of the first passage times.  At the higher temperatures we considered, we find an exponential distribution of these times. At temperatures below $T/T_0 \approx 0.88$ the distribution deviates from an exponential, and increasingly so as temperature is further lowered.  This behavior implies the onset of glassy dynamics\cite{Chandler:2010p6289} with an onset temperature of about 0.88$\,T_0$.  The mean value of these distributions is graphed as $\tau_\mathrm{liq}$ in Fig. \ref{Fi:tau_temp}.

The mean nucleation or mean first-passage time, $\tau_\mathrm{xtl}$, shows expected non-monotonic temperature dependence.\cite{Debenedetti_book} At higher temperatures, nucleation rates increase upon cooling because the barrier to nucleation decreases in size.  In contrast, at lower temperatures, the process of crystallization is slowed by the onset of glassy dynamics. At conditions where the amorphous phase is unstable, $\tau_\mathrm{xtl}$ and $\tau_\mathrm{liq}$ are within two orders of magnitude of each other.  These are average times for crystal nucleation and liquid structural relaxation, respectively.  The distributions of these times are broad, with widths at least as large as the mean values.  Therefore, the distributions of possible times for these respective processes will overlap, which is why a liquid state is no longer physically realizable in this region of the phase diagram.  

The boundary  to unstable amorphous behavior is often referred to as the ``homogeneous nucleation line.'' This terminology is possibly confusing because no significant barrier to nucleation exists in no-man's land.  Indeed, studying the same model with straightforward molecular dynamics in the region of no-man's land, at $T/T_0= 0.72$,  Moore and Molinero conclude that the critical nucleus is less than 10 molecules. \cite{Moore:2010p1923} Coarsening times for relaxing defects in the crystal are necessarily longer than $\tau_\mathrm{xtl}$; and for mW water these times seem to be at least two orders of magnitude larger.\cite{Moore:2010p1923}

The most important point to take from Fig. \ref{Fi:tau_temp} is that time scales for forming crystals are many orders of magnitude larger than those to equilibrate the liquid at standard conditions. This fact explains why straightforward molecular dynamics simulation has thus far proved to be an unreliable probe of phase transitions in supercooled water and related materials. In the case of amorphous phase behavior, close to  or within no-man's land, we see from Fig.~\ref{Fi:tau_temp} that mW water requires time scales to equilibrate that are 3 orders of magnitude larger than those of the normal temperature liquid. Thus, while Moore and Molinero's study of freezing at such conditions~\cite{Moore:2010p1923} is sufficiently long to illustrate likely trajectories leading to a crystal, it is too short to provide quantitative information on the underlying probability distributions that dictate the instability of the liquid phase. It is also not possible from that study to determine if other trajectories exist that might lead to a second metastable liquid phase.

 For real water (or more elaborate atomistic models) equilibration times must account for reorganization that overcomes donor-acceptor asymmetry of hydrogen bonding, a feature that is absent in mW water.  Barriers to orientational reorganization will be comparable to those of translational reorganization. Thus, near the no-man's land boundary one expects equilibration times of, say, ST2 water, to be several orders of magnitude longer than those of mW water. Moore and Molinero~\cite{Moore:2010p1923} estimate it to be seven orders of magnitude longer. This is an issue that can be examined in future studies using methods of rare-event sampling.\cite{Frenkel_book}  For now, however, this paper has demonstrated that time-scale issues do not prohibit the systematic study of reversible phase behavior of water and related systems using the methods of free energy sampling,\cite{Frenkel_book,Binder_book} and such study draws a picture contrary to a widely popularized notion of a second critical point at supercooled conditions.\cite{Poole:1992p2103,Mishima:1998p2948,Mishima:1998p2948,Stokely:2010p6455,Kumar:2005p2095,Fuentevilla:2006p4635,Mishima:2000p4162,Poole:1994p673}  

\begin{acknowledgments} Without implying their agreement with what we write, we are grateful to C. Austen Angell, Sergey Buldyrev, Pablo Debenedetti, Yael Elmatad, Aaron Keys, Valeria Molinero,  Athanassios Panagiotopoulos, Ulf Pedersen, Peter Poole, Eugene Stanley, Patrick Varilly, Benjamin Widom and Michael Widom for helpful discussions regarding our work.  Early work on this project was supported by the Director, Office of Science, Office of Basic Energy Sciences, Materials Sciences and Engineering Division and Chemical Sciences, Geosciences, and Biosciences Division of the U.S. Department of Energy under Contract No. DE-AC02-05CH11231. While for the later stages the authors gratefully acknowledge the Helios Solar Energy Research Center, which is supported by the Director, Office of Science, Office of Basic Energy Sciences of the U.S. Department of Energy under Contract No. DE-AC02-05CH11231.

\end{acknowledgments}

\section*{Supplement}
The prior sections of this paper are published as an article in the \textit{Journal of Chemical Physics}.  Subsequent to our submission of that work and the posting of the first version of this paper, we have received numerous questions about our results for the mST2 model.  Spurred by this interest, this Supplement provides additional information.

\begin{figure}[t]
\begin{center}
\includegraphics[scale=0.4 ]{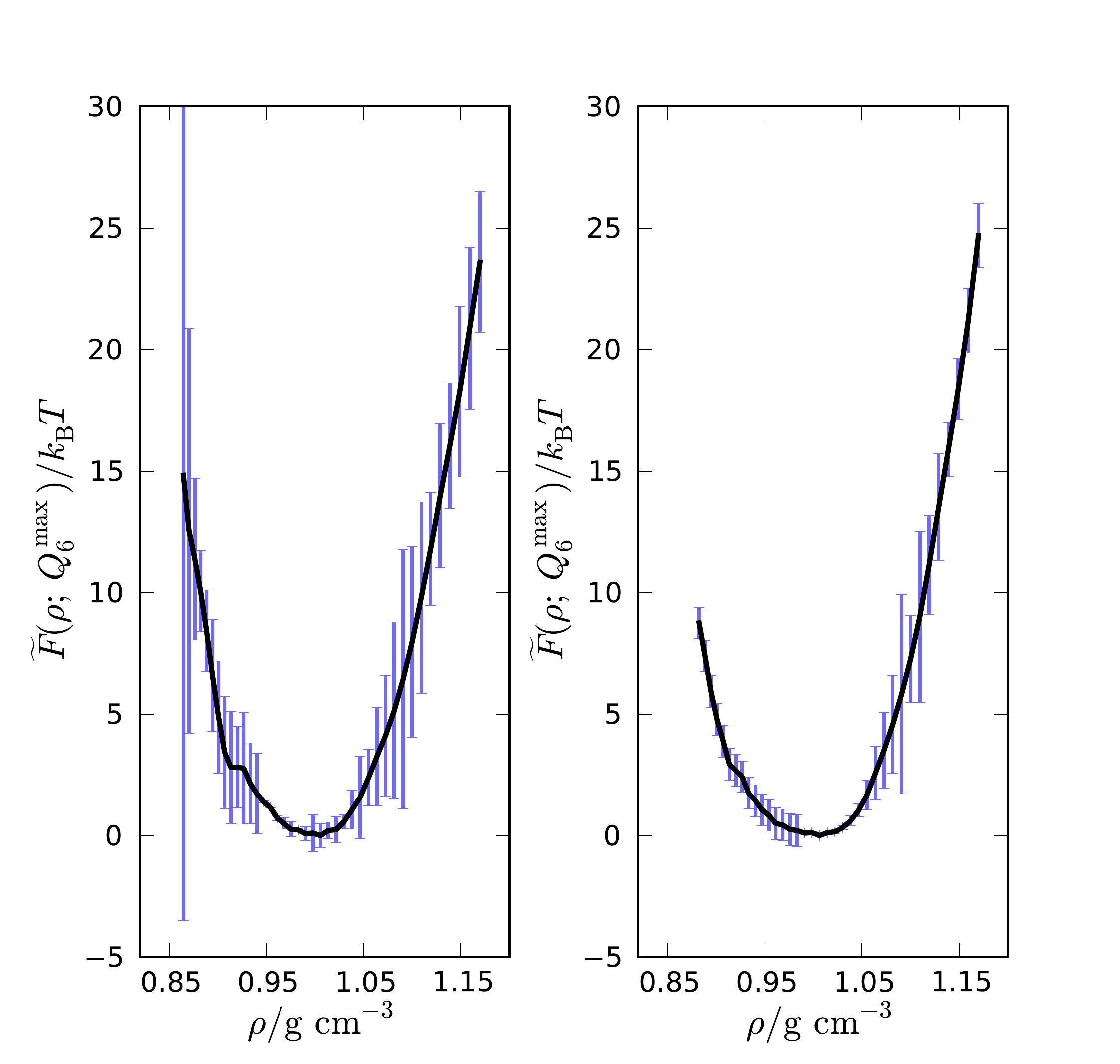}
\caption{The restricted contracted free energy function, $\tilde{F}(\rho; Q_6^{\mathrm{max}})$, for the mST2 model with $N=216$ at the temperature and pressure 235K and 2.2 kbar, respectively, and $\Delta \mu = 0$.  The orientational order parameter $Q_6$ is restricted to $Q_6 \leqslant Q_6^{\mathrm{max}}= 0.13$. The left panel does the integration specified in Eq. (12) with the data presented in Section IV B, specifically Fig. 6.  The right panel does the integration with data obtained with further sampling.  Error bars indicate one standard deviation for the sampled data. The black lines are guides to the eye drawn through the mean values obtained from each sampling bin.
}\label{Fi:contraction_error}
\end{center} 
\end{figure}

The results we present now focus on a contracted free energy function that is restricted to a range of $Q_6$-values, i.e., 
\begin{equation}
e^{-\tilde{F}(\rho; Q_6^{\mathrm{max}})/k_\mathrm{B}T} = \int_0^{Q_6^{\mathrm{max}}} \mathrm{d} Q_6 \,\,e^{-\tilde{F}(\rho, Q_6)/k_\mathrm{B}T}\,.
\end{equation}
Simulation estimates for this function when $Q_6$  is restricted to liquid-basin values are shown in Fig. \ref{Fi:contraction_error}.  The graph on the left, obtained from the same data used to construct Fig. 6, shows statistical uncertainties of the order of $k_\mathrm{B} T$.  The graph on the right, obtained from further sampling, shows uncertainties that are a fraction of that size.  The graph on the left cannot exclude the possibility of a subtle shoulder on the low density side of the density distribution.  Such a shoulder could conceivably lead to a bi-stability at lower pressure, with a barrier between the two basins no larger than $k_\mathrm{B} T$.  This behavior is what has been reported in two simulation studies of the ST2 model.\cite{Liu:2009p776, PooleArXiv}  For this range of densities and $Q_6$, however, the graph on the right excludes the possibility of bi-stability with barrier heights and basin locations reported in Refs. \onlinecite{Liu:2009p776} and \onlinecite{PooleArXiv}.

\begin{figure}[t]
\begin{center}
\includegraphics[scale=0.22 ]{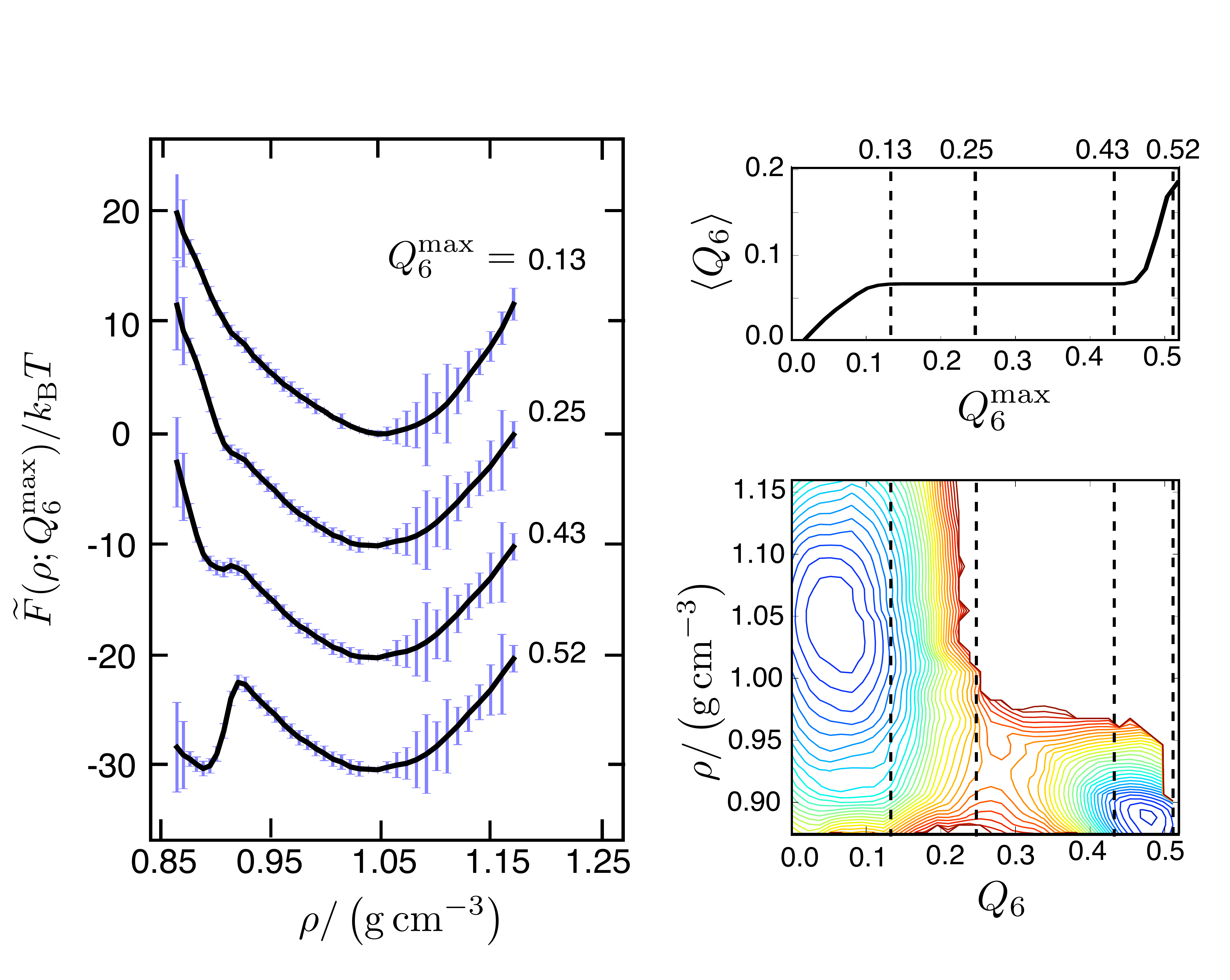}
\caption{To within additive constants, the left graph shows the restricted contracted free energy function, $\tilde{F}(\rho; Q_6^{\mathrm{max}})$, for $N=216$ at several indicated choices of $Q_6^{\mathrm{max}}$.  These functions are obtained from the integration in Eq. (12) with the unrestricted free energy function shown at the bottom right.  The thermodynamic state point for this unrestricted free energy is (235K, 2.2 kbar) and $\Delta \mu = 0.27 k_\mathrm{B}T$. Error bars indicate one standard deviation.  Also shown, upper right, is the mean value of $Q_6$ as a function of the maximum order-parameter value.  Dashed vertical lines in the graphs on the right are at the restricting maximum values for $Q_6$ in the graphs shown on the left.
}\label{Fi:contraction_rho}
\end{center} 
\end{figure}

Dynamics of supercooled water can be glassy, and in that circumstance, only a limited range of $Q_6$ values will be explored by finite-time trajectories without importance sampling.  As such, it is interesting to consider the effect of restricting the range of $Q_6$.  These effects may explain behaviors found by others studying this model.  Figure \ref{Fi:contraction_rho} shows its effect on the contracted free energy of mST2 water.  The particular condition considered is where the free energies of the liquid and crystal basins are equal.  Restricting $Q_6$ to values less than 0.4, however, does not allow the system to reach the crystal basin, and the average of $Q_6$ is an order-of-magnitude smaller.  At the same time, the free energy as a function of density exhibits an inflection or slight minimum.  This feature could be confused with a second liquid basin, but in fact, it is due to the barrier separating liquid from crystal.  The rapid increase in $\langle Q_6 \rangle$ as $Q_6^{\mathrm{max}}$ increases from 0.4 to 0.5 is indicative of the barrier that separates the amorphous and crystal basins. It is only by sampling the full range of $Q_6$ that the free energy shows the phases to be in coexistence and the mean value of $Q_6$ to be indicative of a crystal in a finite simulation system.
 
%\bibliography{ref,ref_papers}

\end{document}